\documentclass[useAMS,usenatbib]{mn2e}
\usepackage{epsfig}
\usepackage{graphicx}
\usepackage{amssymb}
\usepackage{color}

\title[The force-free pulsar magnetosphere]{The spectral simulations of axisymmetric force-free pulsar magnetosphere}
\author[Cao et al. 2015]
{Gang Cao, Li Zhang \thanks{E-mail: lizhang@ynu.edu.cn} and
Sineng Sun
\\
Department of Astronomy, Yunnan University, Key Laboratory of Astroparticle Physics of Yunnan Province, Kunming, 650091, China\\
}

\begin{document}
\pagerange{\pageref{firstpage}--\pageref{lastpage}} \pubyear{2015}

\maketitle

\label{firstpage}

\begin{abstract}
A pseudo-spectral method with an absorbing outer boundary is used to solve a set of the time-dependent force-free equations. In the method, both electric and magnetic fields are expanded in terms of the vector spherical harmonic (VSH) functions in spherical geometry and the divergencelessness of magnetic field is analytically enforced by a projection method. Our simulations show that the Deutsch vacuum solution and the Michel monopole solution can be well reproduced by our pseudo-spectral code. Further the method is used to present the time-dependent simulation of the force-free pulsar magnetosphere for an aligned rotator. The simulations show that the current sheet in the equatorial plane can be resolved well, and the obtained spin-down luminosity in the steady state is in good agreement with the value given by \citet{spi06}.
\end{abstract}


\section{Introduction}

Pulsars are rotating neutron stars with very strong magnetic fields of order $10^{12}$ G. These magnetic fields mediate the conversion of their rotational energy into magnetohydrodynamic (MHD) winds and accelerate particles to very high energies. These particles can produce non-thermal photon emission. High-energy emissions from these objects provide the valuable information about pulsar magnetospheric structure, particle acceleration mechanism, and plasma physics in strong magnetic fields. Since our lack of knowledge of the precise magnetospheric structure, the site of particle acceleration, and microphysical processes, the origin of pulsar emissions still remains uncertain. Recently, significant progresses have been achieved about some aspects of these issues.

An analytical solution of pulsar magnetosphere is only known in vacuum \citep{deu55}. Most of pulsar models generally assume the structure of the pulsar magnetosphere as a retard vacuum dipole magnetic field. However, the vacuum solution would produce an electric field component, $E_{\parallel}$ , parallel to the magnetic field on the the surface of the star. This component of the electric field would pull charges off the surface of the star and initiate pair cascades. The charges created by the pair cascade would short out the accelerating electric field to form a force-free magnetosphere \citep{gol69}. Follow the work of \citet{gol69}, many attempts are made to produce the structure of the pulsar magnetosphere. \citet{sch73} reduced the structure of an axisymmetric magnetosphere to a single equation for the poloidal magnetic flux, which is called as the pulsar equation. Even in the simplest axisymmetric case, the pulsar magnetosphere still remains uncertain for a long time. \citet{mic73} found an exact analytical solution to the pulsar equation in the special case of a magnetic monopole. This solution gives some insights into the far-field region where the field structure is asymptotically radial. \citet{con99} made an significant advance for it. They used a clever iterative algorithm to solve the pulsar equation and found a solution that passes smoothly through the light cylinder. This solution has a region of closed field lines extending to the light cylinder, an open zone with asymptotically monopolar poloidal field lines and an equatorial current sheet which splits at the Y-point into two separatrix surfaces. It was later found that the steady-state solutions with the Y-points exist within the light cylinder \citep{goo04,tim06}. This raises the question to the uniqueness of the CKF solution. The best way of solving this problem is to solve the time-dependent Maxwell equations. Recently, significant progresses have been achieved in modeling of the global structure of pulsar magnetosphere. The time-dependent simulations have been studied by several authors based on different methods \citep{kom06,mck06,spi06,kal09,yu11,par12,pet12}. The time-dependent simulations generally converge to a similar CKF solution and reveal an equatorial current sheet beyond the light cylinder.

In the finite difference schemes, the full expression for the electric current density is not taken into account, the parallel electric current is usually dropped because of the complicated expression. Moreover, it is difficult to impose the exact inner boundary at the stellar surface in the Cartesian coordinate. \citet{par12} improved this situation by implementing a pseudo-spectral method in the spherical coordinate, but the code does not strictly maintain the divergencenessless of the magnetic field at each time step. Recently, \citet{pet12} developed a pseudo-spectral code to improve this situation, the code analytically enforces the divergencenessless of the magnetic field by a projection method. However, the resolutions they used in their simulations are too low to well resolve the current sheet in the equatorial plane, especially for the aligned rotator.

In this work, we focus on the simulation of the axisymmetric pulsar magnetosphere. The time-dependent force-free equations are solved by using the pseudo-spectral method developed by \citet{pet12}. Compared to the method of \citet{pet12}, we reduce the problems from 3-dimensional to 2-dimensional geometry in order to increase the resolution. Moreover, we improve the outer boundary condition by implementing an absorbing boundary layer.
Our simulations show that our pseudo-spectral code can well resolve the current sheet in the equatorial plane by improving the algorithm of \citet{pet12}.

The outline of this paper is as follows. In Section 2 we describe the force-free electrodynamics, and in Section 3 we briefly describe our numerical algorithm. We present the force-free magnetosphere simulation for an aligned rotator in Section 4. Finally, we give our discussion and conclusion in Section 5.

\section{FORCC-FREE ELECTRODYNAMICS}

The time-dependent Maxwell equations are
\begin{eqnarray}
{\partial {\bf B}\over \partial t}&=&-{\bf \nabla} \times {\bf E}\;,\\
{\partial  {\bf E}\over \partial t}&=&{\bf \nabla} \times {\bf B}-{\bf J}\;,
\label{Eq1-2}
\end{eqnarray}
supplemented with two initial condition
\begin{eqnarray}
\nabla\cdot{\bf B}&=&0\;, \\
\nabla\cdot{\bf E}&=&\rho_{\rm e}\;,
\label{Eq3-4}
\end{eqnarray}
where $\rho_{\rm e}=\nabla\cdot{\bf E}$ is the charge density and ${\bf J}$ is the current density. Note that we have set the light speed $c=1$ and $4\pi=1$ throughout this paper.
The force-free approximation implies negligible partial inertial and pressure. Therefore, the Lorentz force acting on a plasma fluid element must
vanish,
\begin{eqnarray}
\rho_{\rm e}{\bf E}+\bf J\times\bf B=0\;,
\label{Eq5}
\end{eqnarray}
which implies the force-free condition $\bf E\cdot \bf B=0$.
From the force-free condition and the Maxwell equations, the current sheet is uniquely determined to be \citep{gru99,bla02}
\begin{eqnarray}
{\bf J}= \rho_{\rm e} {{\bf E} \times {\bf B} \over B^2}+
{({\bf B}\cdot {\bf \nabla}\times {\bf B}-{\bf E}\cdot {\bf \nabla}\times {\bf E}){\bf{B}}
\over B^2}\;.
\label{Eq6}
\end{eqnarray}

The two terms in the right side of Eq. (\ref{Eq6}) have following physical meaning: the first term is the drift current perpendicular to $\bf B$, and the second term is conduction current parallel to  $\bf B$, which maintains the force-free condition. Note that the full electric current expression is not taken into account in the finite-difference schemes. The field-aligned component is usually dropped because of the intricate expression including spatial derivatives. It is tricky to handle with finite-difference schemes, which needs the interpolation of both field and field derivatives. However, it is easy for the spectral method to handle with spatial derivatives. Therefore, we include the full electric current expression in our pseudo-spectral code.

\section{PSEUDO-SPECTRAL ALGORITHM}

We now briefly describe the pseudo-spectral algorithm introduced by \cite{pet12} and present an improvement for the algorithm of \cite{pet12} here. The main ingredients are the VSH expansion of the electric and magnetic fields, the force-free condition and the divergencelessness constraints on {\bf B}, an exact enforcement of boundary condition, an explicit time integration with the third-order Adam-Bashforth scheme, and a spectral filtering.

In the pseudo-spectral algorithm, the electric and magnetic fields are expanded in terms of the VSHs by the expressions \citep[e.g.,][]{pet12}
\begin{eqnarray}
  \label{eq:B_VSH}
  \mathbf{B} & = & \sum_{l=0}^\infty\sum_{m=-l}^l
  \left(B^r_{lm} \mathbf{Y}_{lm} + B^{(1)}_{lm} \mathbf{\Psi}_{lm}+
    B^{(2)}_{lm}\mathbf{\Phi}_{lm}\right)\;,\\
  \label{eq:E_VSH}
  \mathbf{E} & = & \sum_{l=0}^\infty\sum_{m=-l}^l
  \left(E^r_{lm} \mathbf{Y}_{lm} + E^{(1)}_{lm}\mathbf{\Psi}_{lm}+
    E^{(2)}_{lm}\mathbf{\Phi}_{lm}\right)\;.
\end{eqnarray}
From the expansion coefficients, we can easily compute the linear differential operators like $\nabla\cdot$ and $\nabla\times$ in the coefficient space and then transformed back to the real space to advance the solution in time. For a detail discussion about the definitions and  properties of VSHs, see \cite{pet12}. In the radial coordinate, the spectral coefficients are expanded into the Chebyshev function such that
\begin{eqnarray}
B^r_{lm}(t,r)=\sum_{k=0}^{N_{r}-1}B^r_{klm}(t) T_{k}(r)\;.
\end{eqnarray}
The radial derivative can be easily computed by the three-term recurrence relation, see \citet{can06}.

\begin{figure*}
\begin{tabular}{cccccc}
  \includegraphics[width=6.cm,height=8cm]{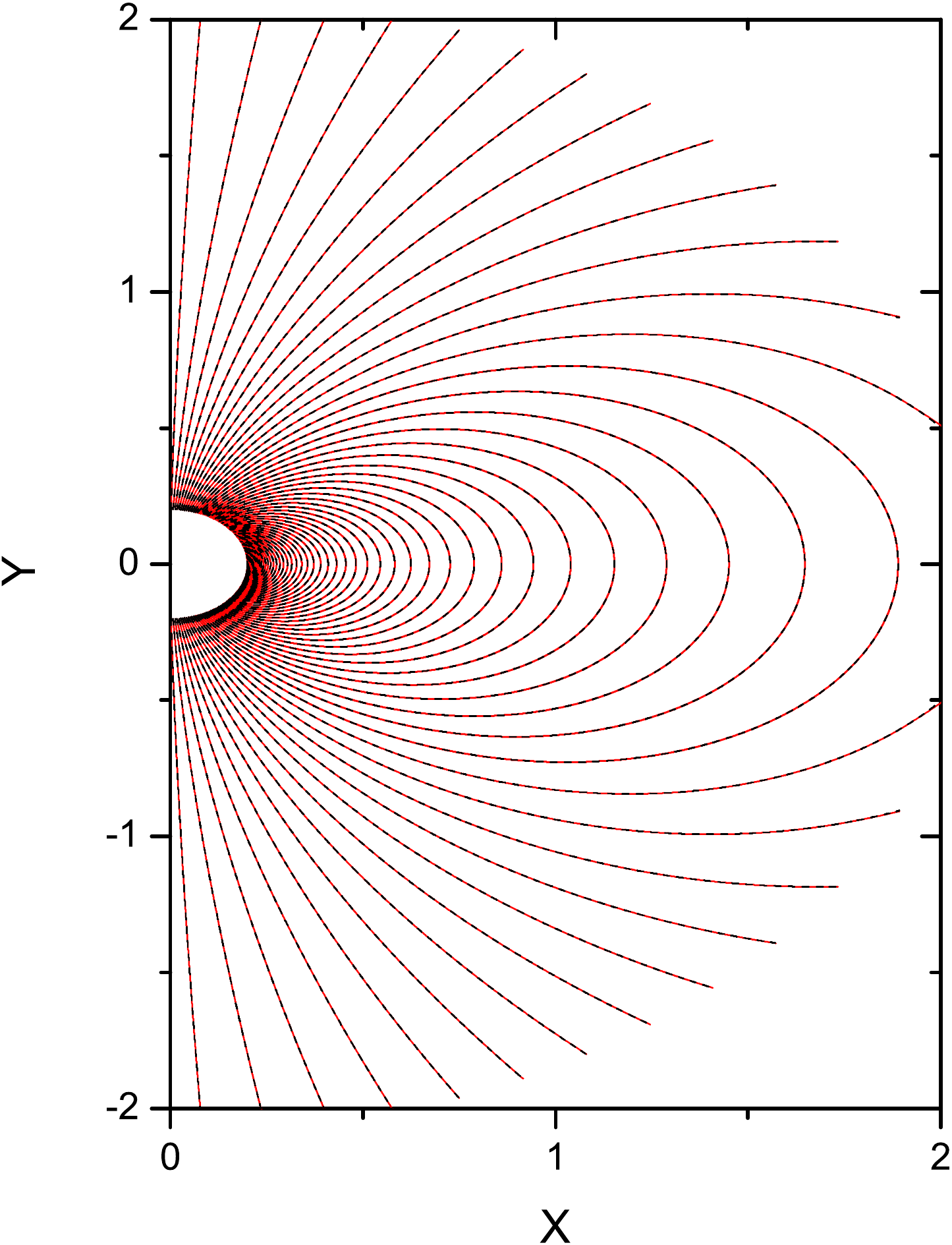} \qquad\qquad
  \includegraphics[width=6.cm,height=8cm]{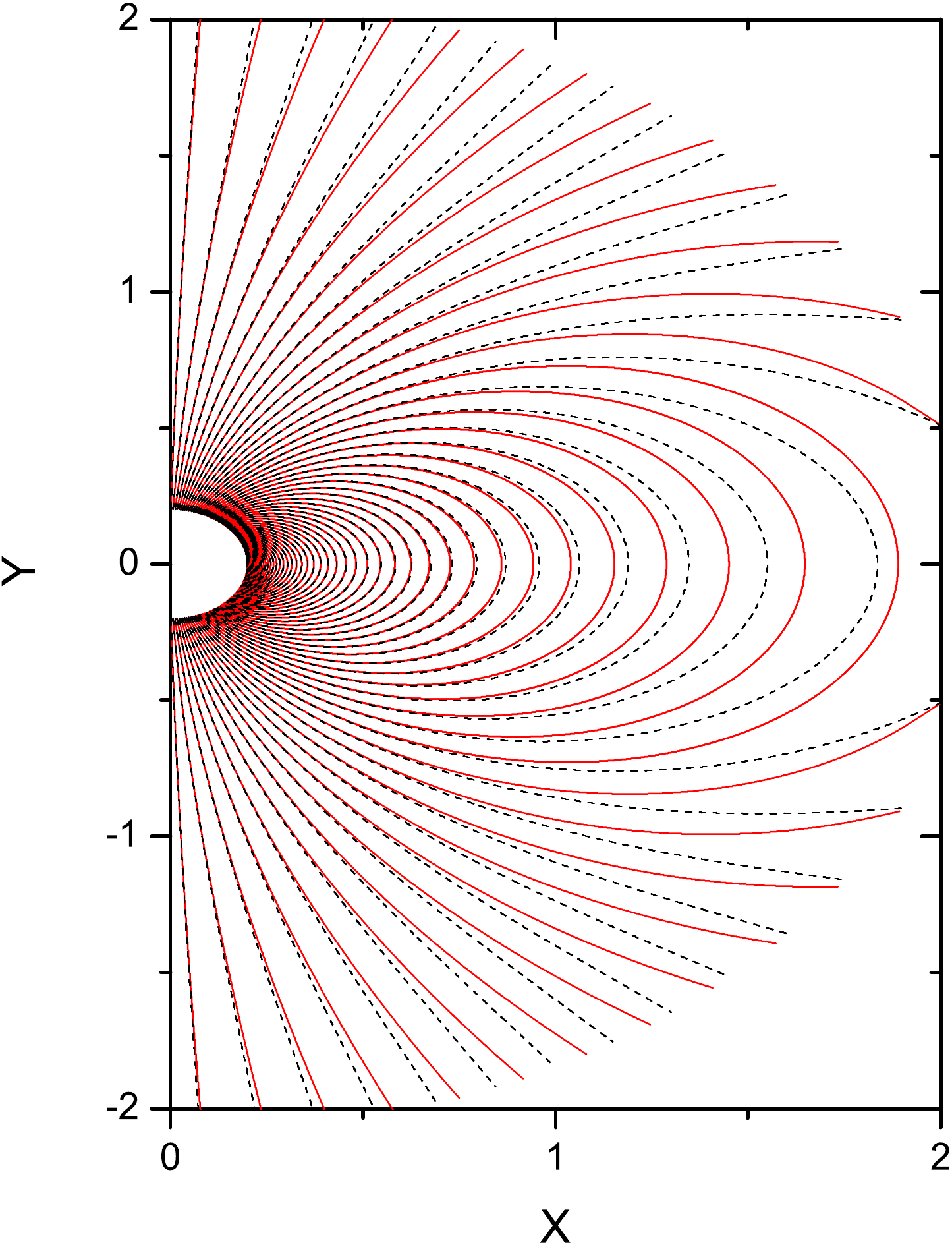}
\end{tabular}
\caption{Magnetic field lines of the aligned Deutsch field in the poloidal plane. The numerical and analytical solutions are shown as the black dashed and red solid lines, respectively. Left panel: the numerical solution with an absorbing boundary layer. Right panel: the numerical solution without an absorbing boundary layer.}
\end{figure*}

During the time evolution of electromagnetic field, the initial force-free configuration may develop some regions in which the force-free condition is violated. Therefore, we enforce the $\bf E \cdot \bf B=0$ and $ E<B$ conditions regularly. At each time step, we subtract the parallel electric field by adjusting the electric field as ${\bf {E}}= {\bf {E}}-{{ {\bf {E}} \cdot \bf {B} } \over {B^2}} \bf {B}$. If this new electric field does not satisfy the condition $ E<B$, we reset the electric field as $E=B$, such that ${\bf {E}} = {\bf {E}} \sqrt{{B^2}\over{E^2}}$. In the force-free equation, the magnetic field is divergencelessness ($\nabla \cdot B=0$). This property can be analytically enforced by the expansion below:
\begin{equation}
{\bf {B}} = \sum_{l=1}^\infty\sum_{m=-l}^l
 \left(\nabla \times [f_{lm}^{B}(r,t) \, { {\bf{\Phi}}_{lm} }] + g_{lm}^{B}(r,t) \, {\bf{\Phi}}_{lm} \right),
\end{equation}
where $\{f^B_{lm}(r,t),g^B_{lm}(r,t)\}$ is the expansion coefficients of $\bf {B}$. This expansion is divergencelessness by definition. Therefore, to impose the divergencelessness constrain, the magnetic field is projected onto the subspace defined by equation (10) at each time step. Specifically, we perform a forward transform from the vector $\bf B$ to the coefficients $\{f^B_{lm}(r,t),g^B_{lm}(r,t)\}$  and a backward transform from the coefficients to the vector $\bf B$.

We follow the method of \citet{pet12} to impose the exact inner boundary condition. However, it is difficult to exactly handle with the outer boundary condition on a sphere of a finite radius. In fact, the approximate boundary treatment described in \citet{pet12} cannot well prevent the reflect waves from the outer boundary.
Therefore, we improve the outer boundary condition by implementing an absorbing boundary layer, which can efficiently avoid the reflection from the outer boundary, see \citet{par12} for a description of the absorbing boundary layer. After the spectral transform from the real space to spectral space is performed, a set of partial differential equations are replaced by a larger set of ordinary differential equations with appropriate initial and boundary conditions. We use a third-order Adam-Bashforth scheme to advance the solution at each time step. In order to ensure the long-time stability of the algorithm and increase the convergent rate of the solution, it is necessary to filter the high-frequency mode to reduce the aliasing errors and the Gibbs phenomenon. The spectral filtering is performed at each time step. We use an eighth-order exponential filter in all directions by the expression
\begin{equation}
\sigma(\eta) = \textrm{e}^{-\alpha\,\eta^\beta}\;,
\end{equation}
where $\eta$ ranges from 0 to 1 and the parameter $\alpha\geq0$. For instance, in the radial direction, the filtering spectral coefficient is given as
\begin{equation}
B^r_{lm}(t,r)=\sum_{k=0}^{N_{r}-1}\sigma(\eta)B^r_{klm}(t) T_{k}(r)\;,
\end{equation}
where $\eta=k/(N_r-1)$ for $k\in[0,1,...,N_r-1]$.

\section{RESULT}

In the force-free electrodynamics, discontinuity in the magnetic field appears in the equatorial current sheet. It is necessary for a large number of collocate points to capture the discontinuity in the current sheet. In the paper by \citet{pet12}, the equatorial current sheet is not well resolved because of the very low resolution, especially for the aligned rotator.
Therefore, in order to increase the resolution and capture the current sheet, we reduce the 3D problem to 2D one by assuming axisymmetry. Specially, we set the spherical harmonics index to zero, $m=0$, in our simulation. Before starting to simulate the force-free magnetosphere for an aligned rotator, we test our 2D code against some known analytical solutions, which include the Deutsch vacuum field solution for the aligned rotator and the force-free monopole solution.

In our simulations, we adopt the following normalization: $B_{\rm L}=\Omega=\mu=c=1$, where $B_{\rm L}$ is the magnetic field strength at the light cylinder in the equatorial plane, $\Omega$ is the stellar angular velocity, $\mu$ is the magnetic dipolar monument, and $c$ is the light velocity. Therefore, the light cylinder radial is $r_{\rm L}=1$.

\subsection{The Deutsch solution for an aligned rotator }

The Maxwell equations have an exact analytical solution in the vacuum case, which is derived by \citet{deu55}. The expressions for the Deutsch vacuum solution can be found in \citet{deu55} and \citet{pet12}.
We start our simulation with an aligned dipole magnetic field and zero electric field outside the star. In the stellar surface, we enforce the inner boundary condition with a corotating electric field ${\bf {E}} = -( {\bf \Omega } \times \bf r ) \times \bf B$. The approximate outgoing boundary cannot work well for the vacuum case \citep{pet14,pet15}. Therefore, we use an absorbing boundary layer to avoid the reflection from the outer boundary. We set the computational domain to be $r\in (0.2 - 3)$ and add an absorbing boundary layer at $r=2$ next to the outer boundary. A resolution of $N_r \times N_{\theta} \times N_{\phi}=64 \times 4 \times 1$
is sufficient to get a good accurate solution inside the absorbing layer. We also try the higher resolution with $N_r \times N_{\theta} \times N_{\phi}=128 \times 8 \times 1$, no significant improvement is found in the computed solution.

We let the system evolve for about 4 rotational period, $t=8\pi$. The final magnetic field lines in the poloidal plane are shown as the black dashed curves in the left panel of Figure 1. The analytical solution is also shown as the red solid curves. The numerical solution is in good agrement with the analytical solution, and they are hardly distinguishable. For comparison, the numerical solution in which we do not add an absorbing boundary layer is also shown in the right panel of Figure 1. It can be seen that the numerical solution strongly deviates from the analytical solution near the outer boundary. The test shows that the absorbing boundary condition can efficiently avoid the reflection from the outer boundary and make the system reach a nearly steady-state configuration. The simulation demonstrates that our pseudo-spectral code can compute accurately vacuum electromagnetic fields.

\begin{figure}
\centerline{\epsfig{file=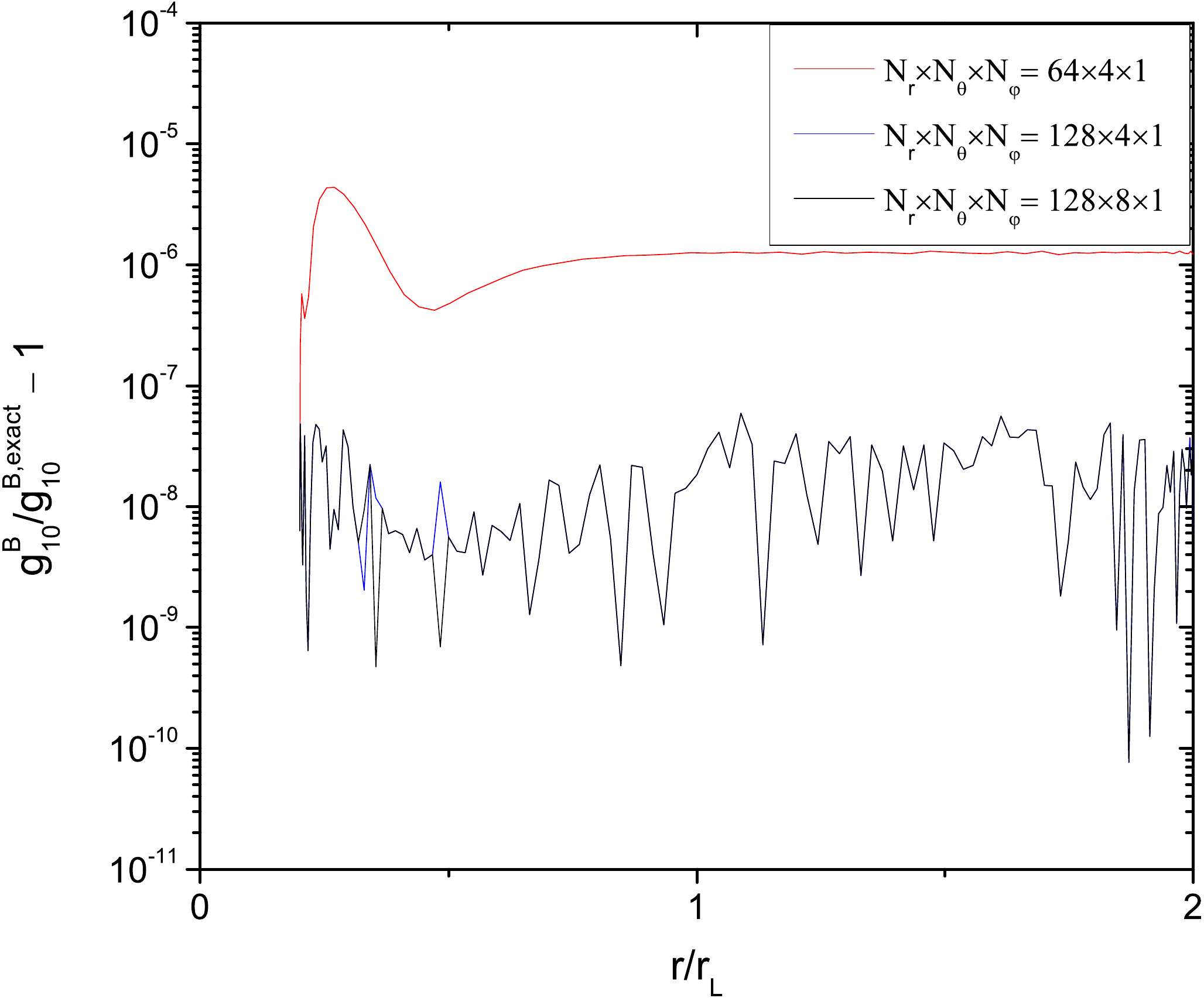, width=8cm}}
\caption{ The relative errors of the function $g_{10}^{B}(r)$ for the force-free monopole with different resolutions.
\label{fig2}}
\end{figure}

\subsection{The force-free monopole solution }

The force-free monopole solution is given by \citep{mic73}
\begin{equation}
{\bf{B}} = B_{\rm L} \frac{r_{\rm L}^2}{r^2} {\bf e}_{r} - B_{\rm L} \frac{r_{\rm L}}{r} \sin {\theta} {\bf e}_{\phi}\;.
\end{equation}
This magnetic field can be expressed as a VSH expansion,
\begin{equation}
{\bf{B}} = B_{\rm L} \frac{r_{\rm L}^2}{r^2} {\bf e}_{r} + g_{10}^{B,\rm exact}(r) {\bf {\Phi}}_{10}\;,
\end{equation}
where
\begin{equation}
g_{10}^{B,\rm exact}(r)=\sqrt{\frac{8\pi}{3}} B_{\rm L} \frac{r_{\rm L}}{r}\;.
\end{equation}

We start the simulation with a pure monopolar magnetic field and a zero electric field outside the star, as before. During the time evolution of the electromagnetic field, we keep the component $B_r$ constant in time.
We only need to test the spectral coefficient $g_{10}^{B,\rm exact}(r)$ in the $B_{\phi}$ component. We do not add an absorbing layer, since the monopole field lines are radial outgoing waves, for which the approximate boundary treatment is very effective. In our simulation, the computational domain is set to be $r\in (0.2 - 2)$. We find a resolution of $N_r \times N_{\theta} \times N_{\phi}=64 \times 4 \times 1$ is sufficient to get a high accurate solution. In order to prove the accuracy of our code, we compare the rate $g_{10}^{B}(r)/g_{10}^{B,\rm exact}(r)$ with unity, which is depicted in Figure 2. Contrary to the Deutsch vacuum field, the force-free solution does not suffer from the effect of the outer boundary. The numerical solution is in good consistency with the exact solution with high accuracy in the whole computational domain, even though in the low resolution case. After increasing the higher resolutions in the $r$ and $\theta$ directions, we find that the errors rapidly decrease with the increase resolution in the $r$ direction, but no significant improvement when increasing the resolution in the $\theta$ direction (see Figure 2).

\begin{figure*}
\begin{tabular}{cccccc}
  \includegraphics[width=5.cm,height=7cm]{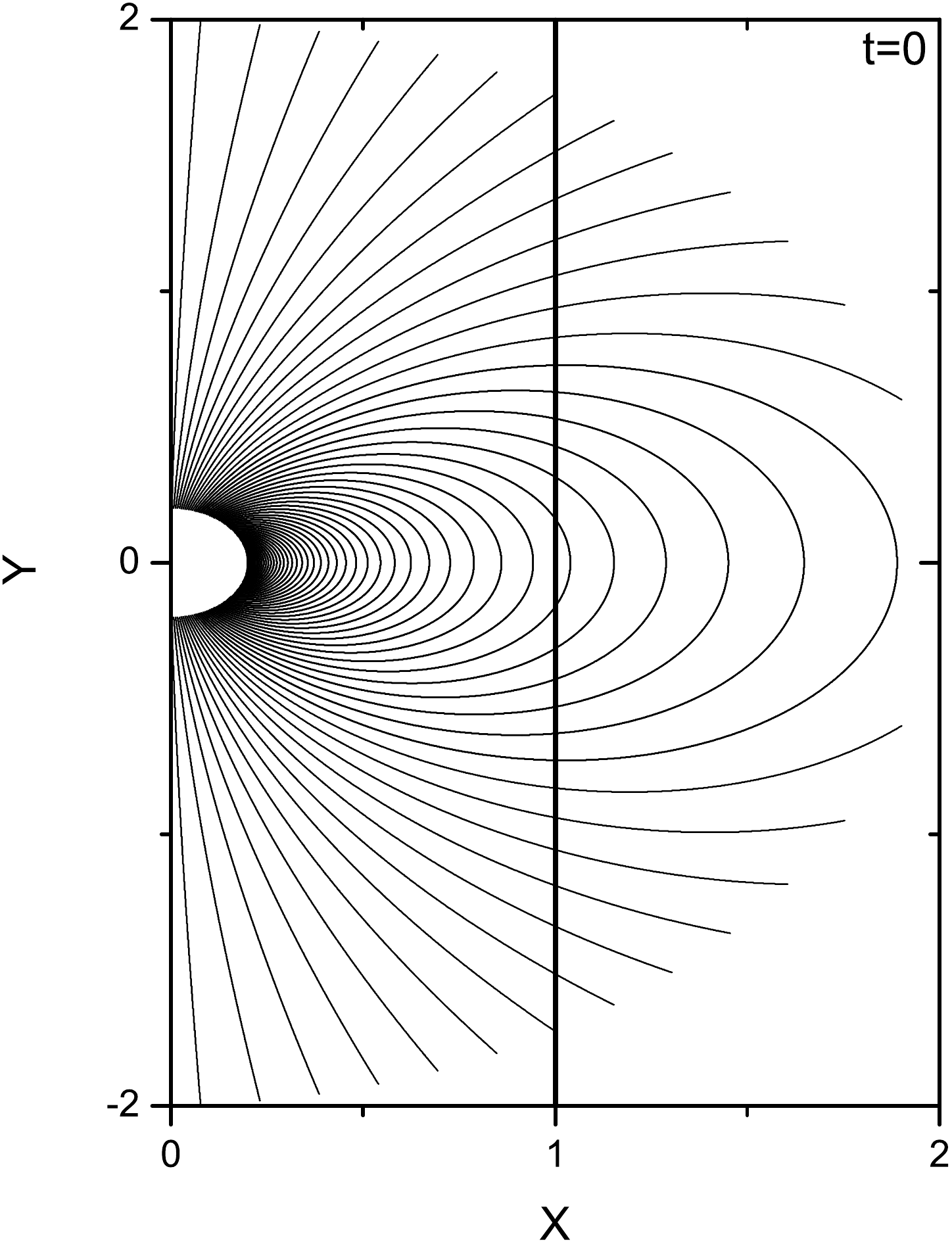} \qquad
  \includegraphics[width=5.cm,height=7cm]{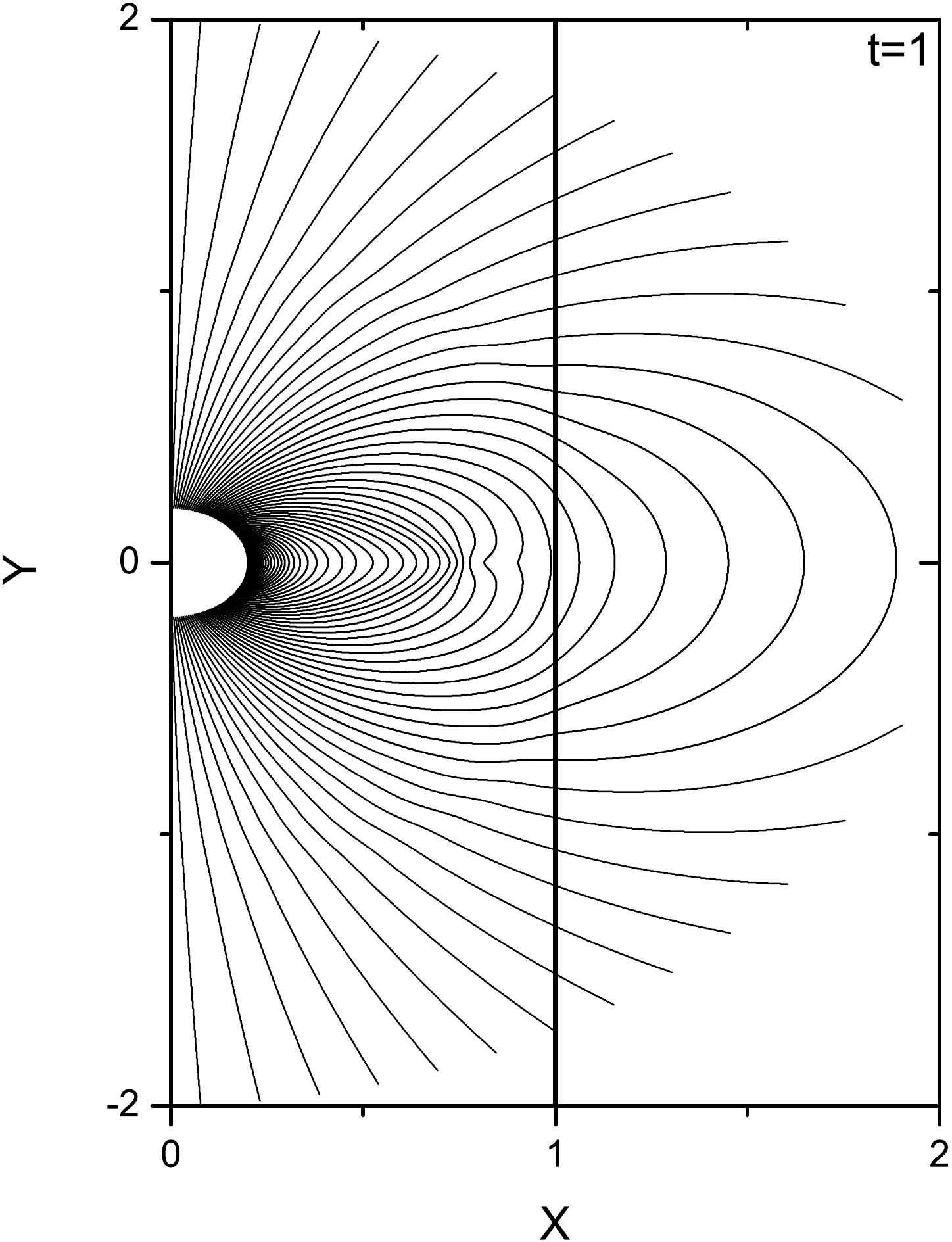} \qquad
  \includegraphics[width=5.cm,height=7cm]{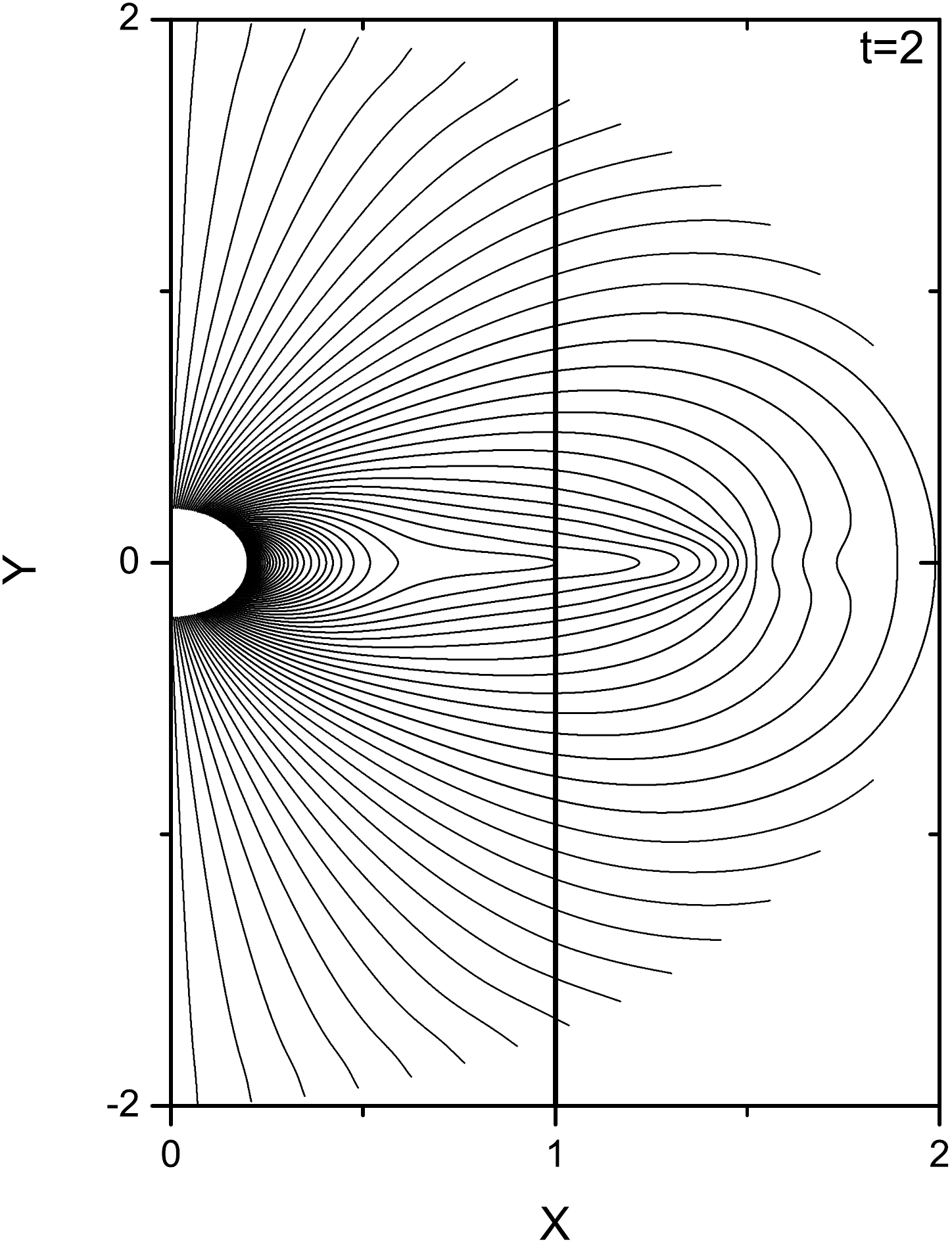} \\
  \includegraphics[width=5.cm,height=7cm]{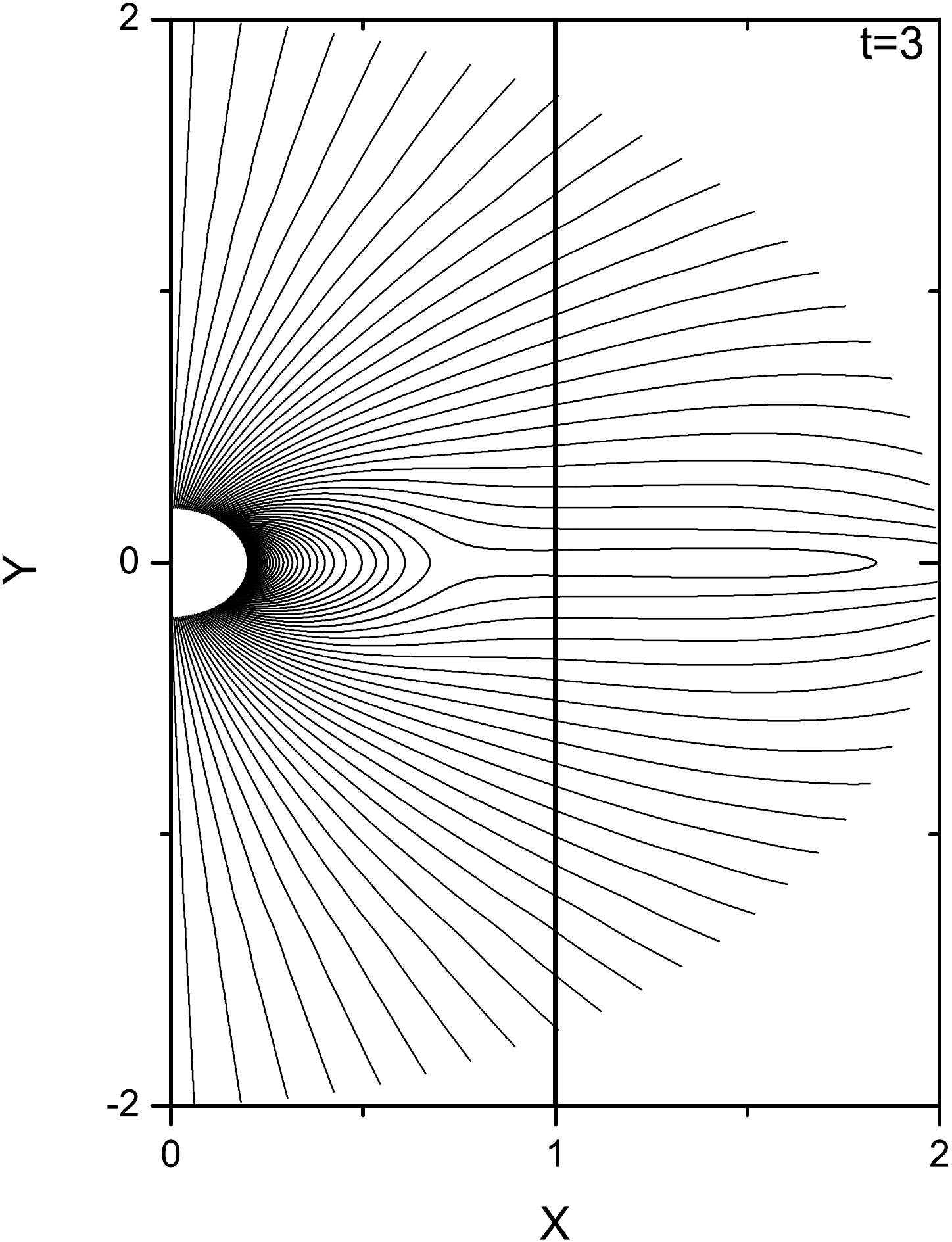} \qquad
  \includegraphics[width=5.cm,height=7cm]{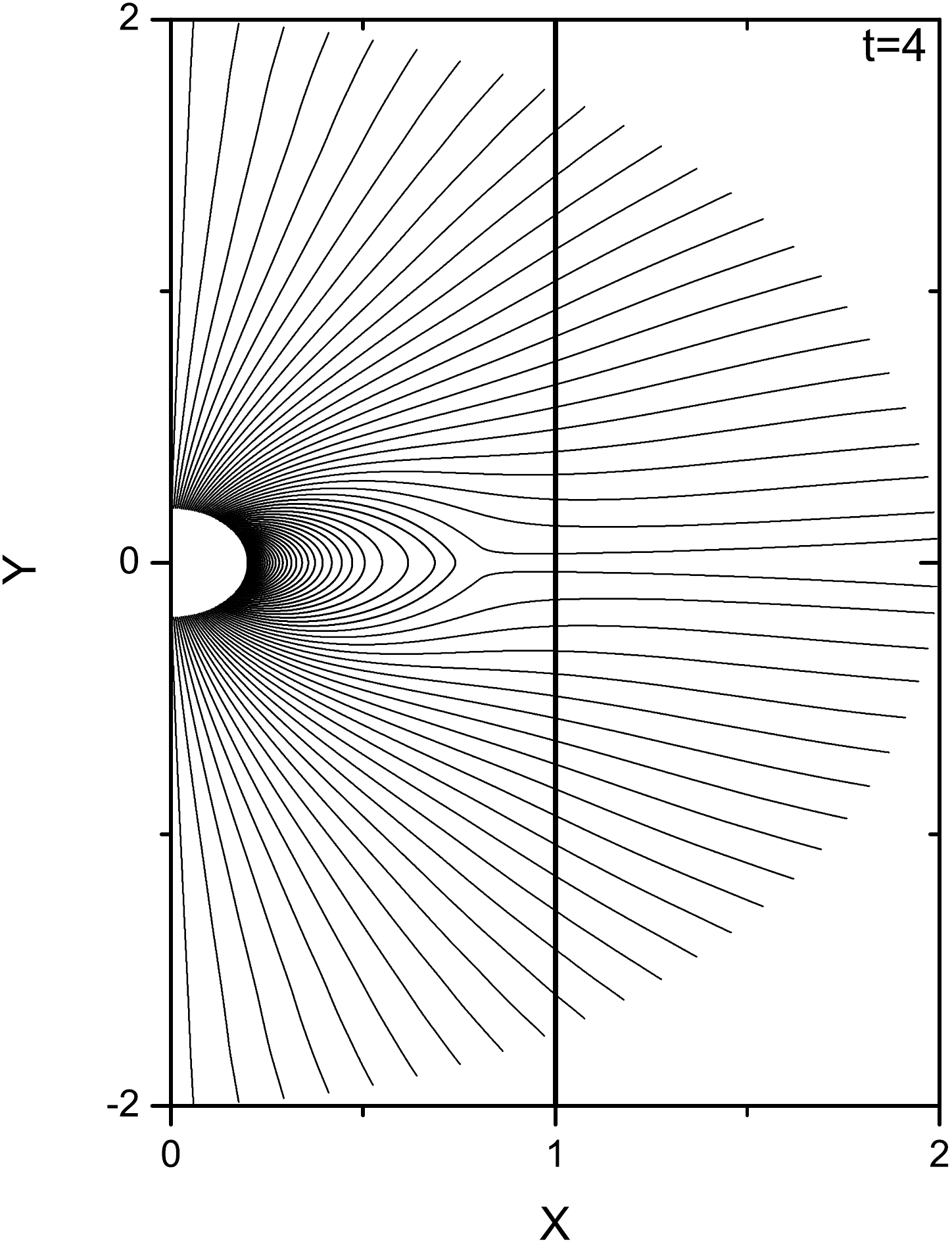} \qquad
  \includegraphics[width=5.cm,height=7cm]{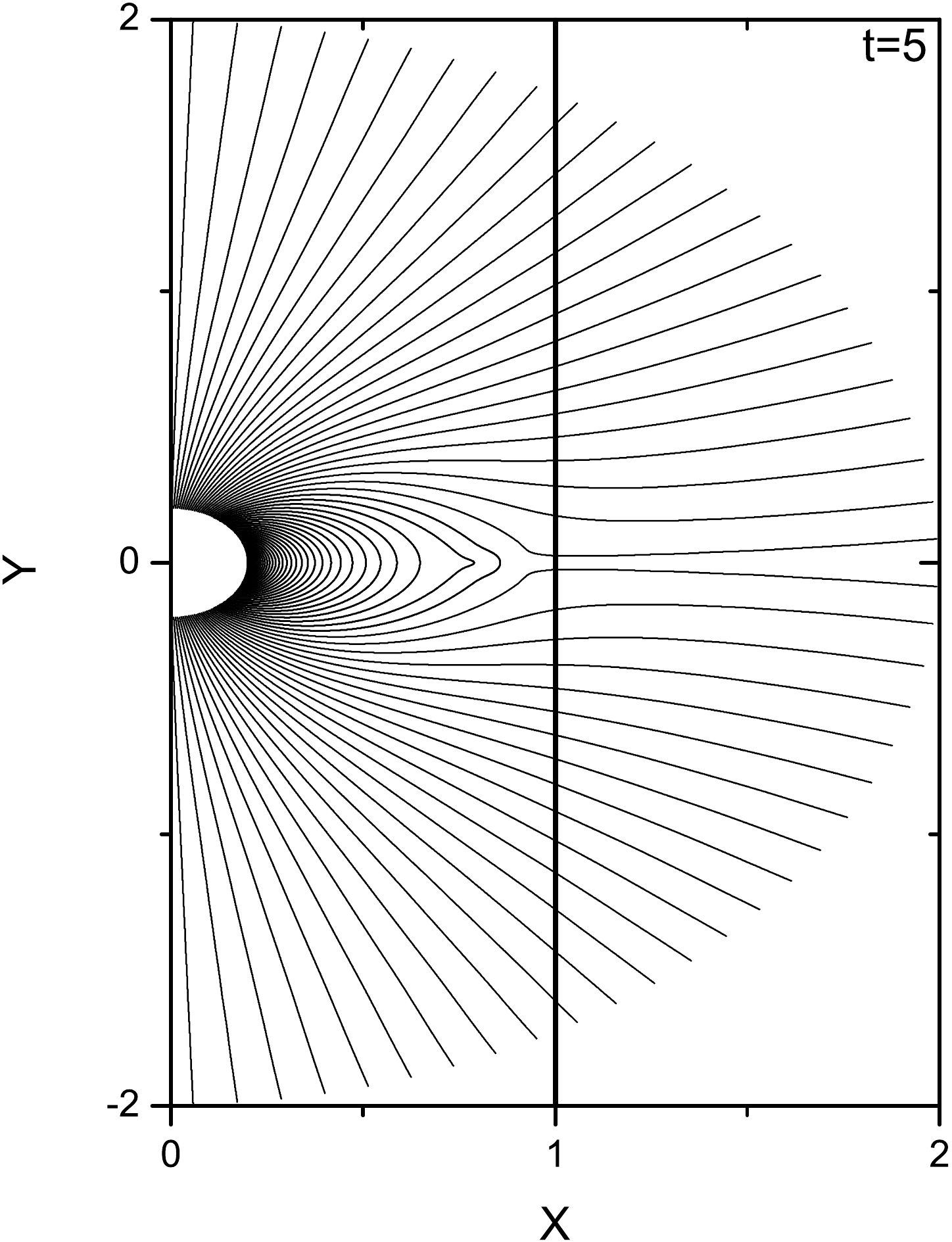} \\
  \includegraphics[width=5.cm,height=7cm]{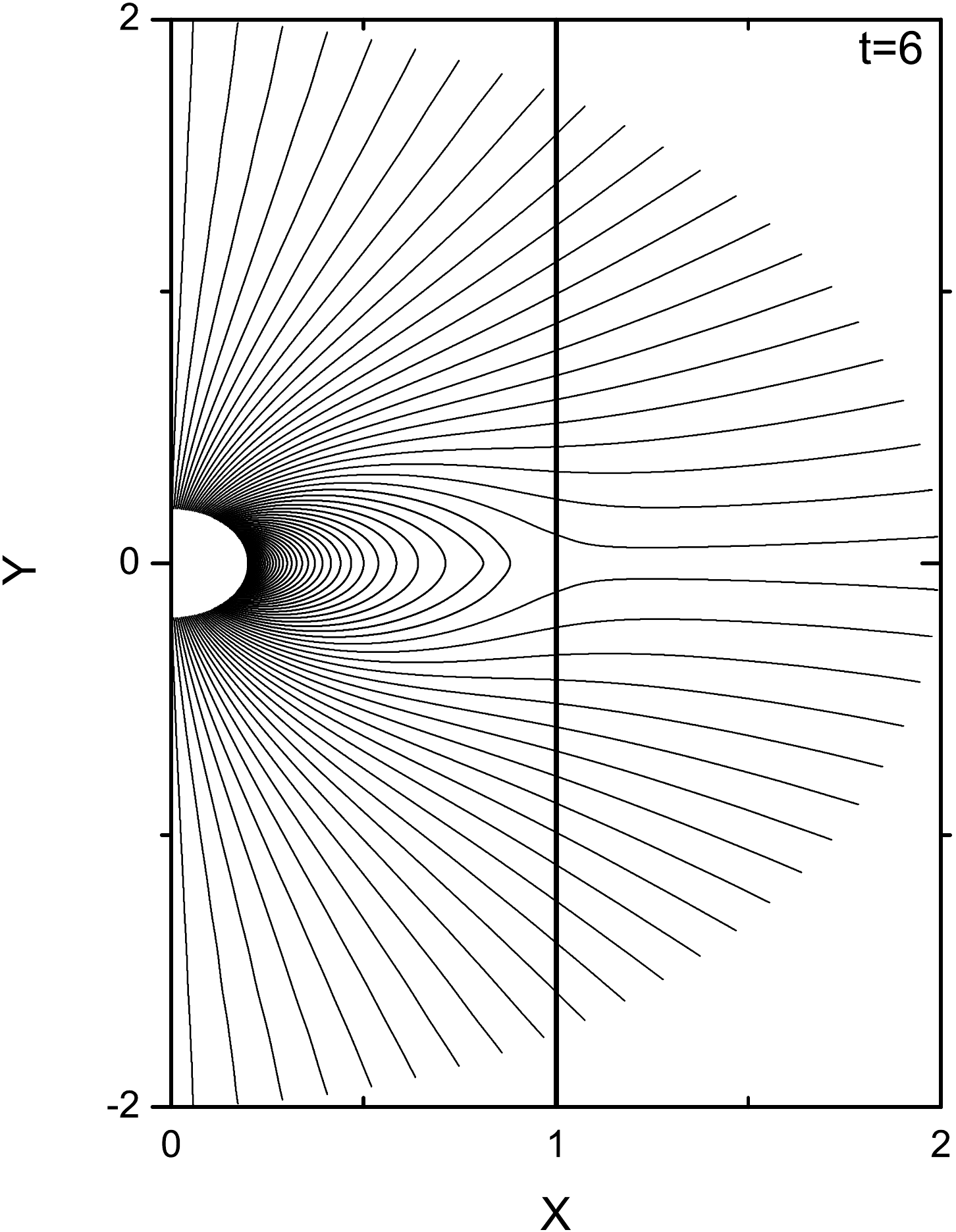} \qquad
  \includegraphics[width=5.cm,height=7cm]{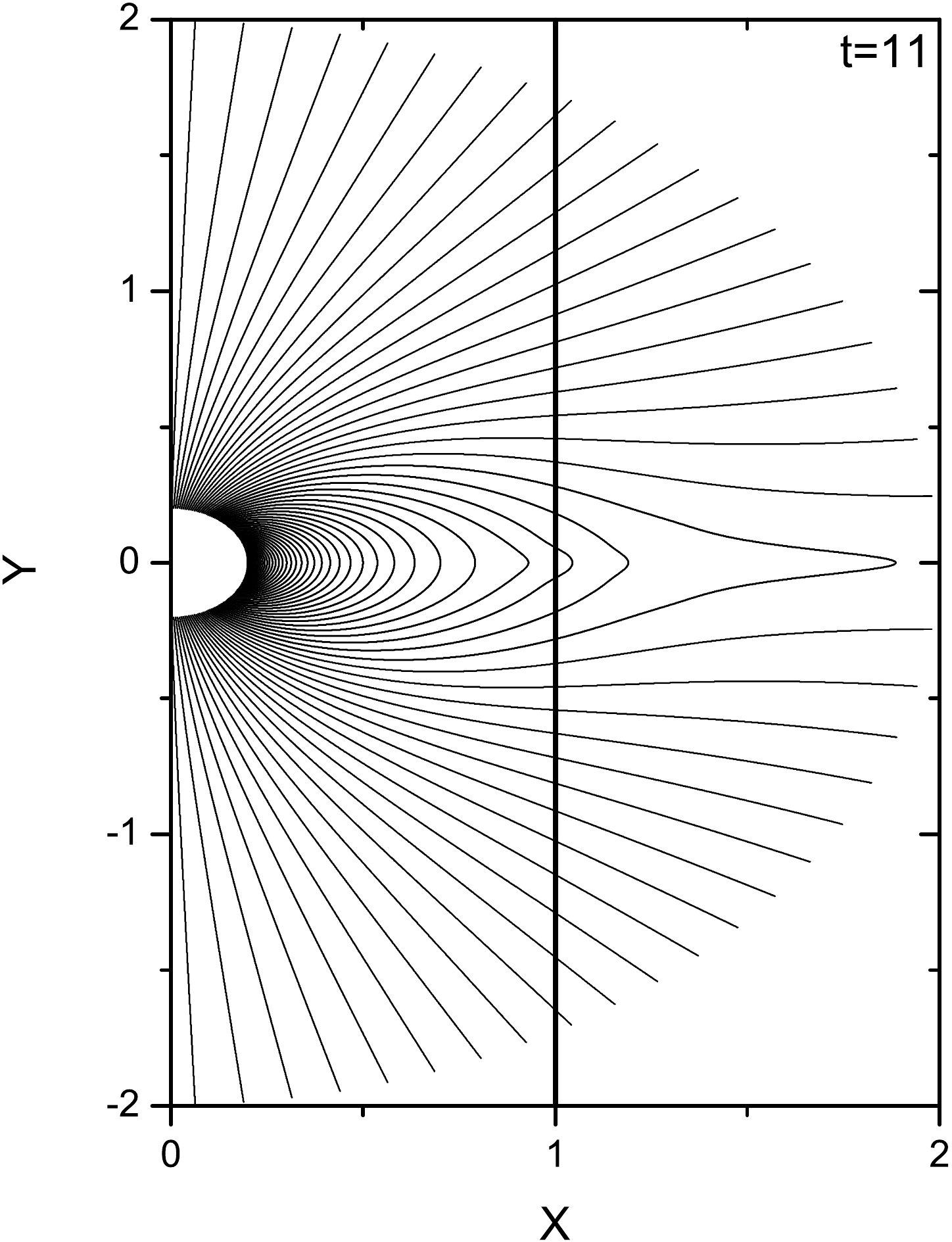} \qquad
  \includegraphics[width=5.cm,height=7cm]{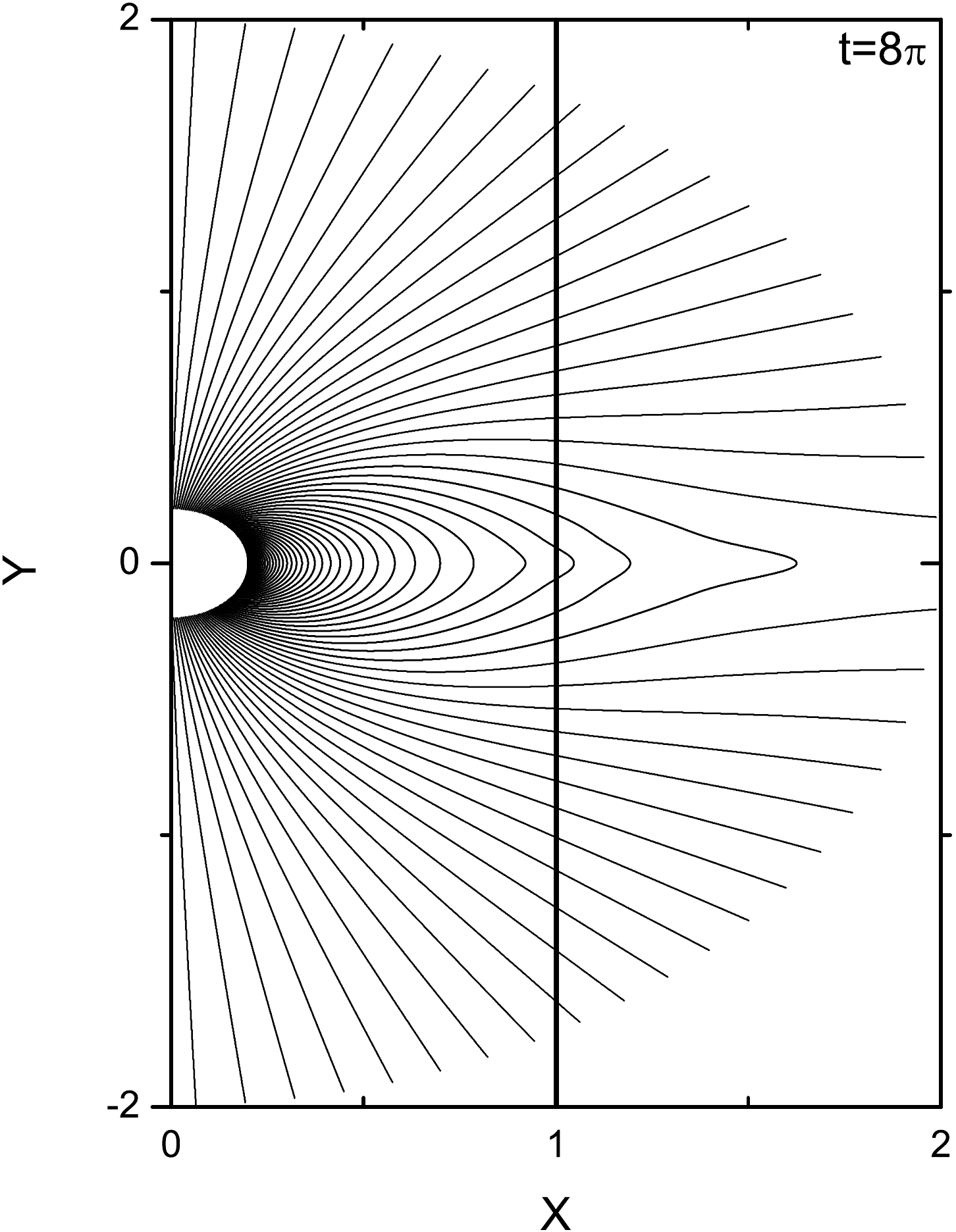}
\end{tabular}
\caption{Time sequence of poloidal field lines for an aligned rotator. Light cylinder is shown as the thick  vertical line. }
\end{figure*}

\subsection{The force-free solution for an aligned rotator}

After testing the code against the analytical solutions, we present the force-free magnetosphere simulation for an aligned rotator by using our pseudo-spectral code. In our simulations, we set the computational domain to be $r\in (0.2 - 3)$ and use an absorbing boundary layer starting at $r=2$ and extending to the outer boundary $r=3$. In order to capture the current sheet, we use the super spectral viscosity (SSV) filter in the radial direction, which is explained in depth in \citet{boy98}. We perform the simulations with different resolutions, a resolution with $N_r \times N_{\theta} \times N_{\phi}=128 \times 181 \times 1$ is adopted in our simulation, the behavior is similar for all other resolutions.

The magnetic field is initialized to be an aligned dipolar $B_r=2\cos{\theta}/r^3$, $B_{\theta}=\sin{\theta}/r^3$, $B_{\phi}=0$. The electric field is set to be zero, ${\bf E}=0$, except for the stellar surface where we enforce the inner boundary condition with a corotating electric field.

In Figure 3, we show the time sequence of the poloidal field lines to approach the steady state. An Alfv\'{e}n wave is launched from the stellar surface, the wave transports the charges and currents outwards, and the associated electric field sets the field lines in rotation. As the waves from two opposite hemisphere meet in the equatorial palne, the closed zone extends outwards. The field lines are pulled out by the electromagnetic pressure, the closed field lines are on the way to opening. As the more field lines are opened, the magnetic field in the equatorial plane is decreased. After the half of the rotation period, the magnetic field reaches to zero. Beyond this point, the system forms a current sheet in the equatorial plane with a opposite magnetic field, and the field lines form a closed-open configuration with the Y-point inside the light cylinder. The Y-point moves slowly towards to the light cylinder, as the some of the open field lines reconnect in the current sheet. The Y-point reaches the light cylinder at $t\sim5.5$, the magnetosphere forms a closed-open CKF configuration, which is shown in Figure 4. During the following evolution, the field lines continue to reconnect in the current sheet outside the light cylinder until the system reaches to the steady state. After about 2.5 rotation period, the solution approaches to the steady state. In our steady-state solution, some field lines are closed outside the light cylinder, which is due to the numerical dissipation caused by spectral filtering. Our solution is similar to that of \citet{par12}.

The radial current density in the steady-state solution, normalized to the Goldreich-Julian (GJ) current density $J_{\rm GJ}=-\frac{\Omega  \cdot {\bf B}}{2\pi}$, is shown in Figure 5. The electric current flows out along the open field lines and returns to the surface of the star in the current sheet, and the radial current density is smaller than the GJ current density. Our simulation result is agreement with the state-of-the-art steady-state solution of \citet{tim06}.

The Poynting flux integrated over a sphere of radial distance $r$ is given by
\begin{equation}
L=\int r^2 d\Omega \,{\bf S \cdot}{\bf e}_{r}\;,
\end{equation}
where $ {\bf {S}}= {\bf {E \times B}}/4\pi$ is the Poynting vector, $d \Omega$ is the infinitesimal solid angle, and $\Omega$ is the full sky angle of $4\pi$ sr.

The Poynting flux for the force-free aligned rotator is given by \citep{spi06}
\begin{equation}
L_{\rm aligned}=\frac{B_{\star}^2 R_{\star}^6 \Omega_{\star}^4}{c^3}\;.
\end{equation}
The normalized Poynting flux $L/L_{\rm aligned}$ across the sphere of radial $r$ is shown in Figure 6. The Poynting flux inside the light cylinder remains constant and well agrees with the theoretic value given by \citet{spi06}. However, the numerical dissipation due to spectral filtering slightly changes the energy conservation outside the light cylinder. Note that the Poynting flux given by \citet{pet12} is much larger than the theoretic value. Therefore, our result is much better than that of \citet{pet12} because of the higher resolution in our simulation.
The open magnetic flux through the light cylinder is $\Psi_{\rm open}=1.40 \, \Psi_{\rm dip}$, where $\Psi_{\rm dip}=\frac{B_{\star} R_{\star}^3 \Omega_{\star}}{c}$ is the open magnetic flux for an aligned vacuum dipole.  This is larger than the value $\Psi_{\rm open}=1.23 \, \Psi_{\rm dip}$ given by \citet{tim06} and \citet{con05}, because some closed lines pass through the light cylinder due to artificial dissipation by spectral filtering.

\begin{figure}
\centerline{\epsfig{file=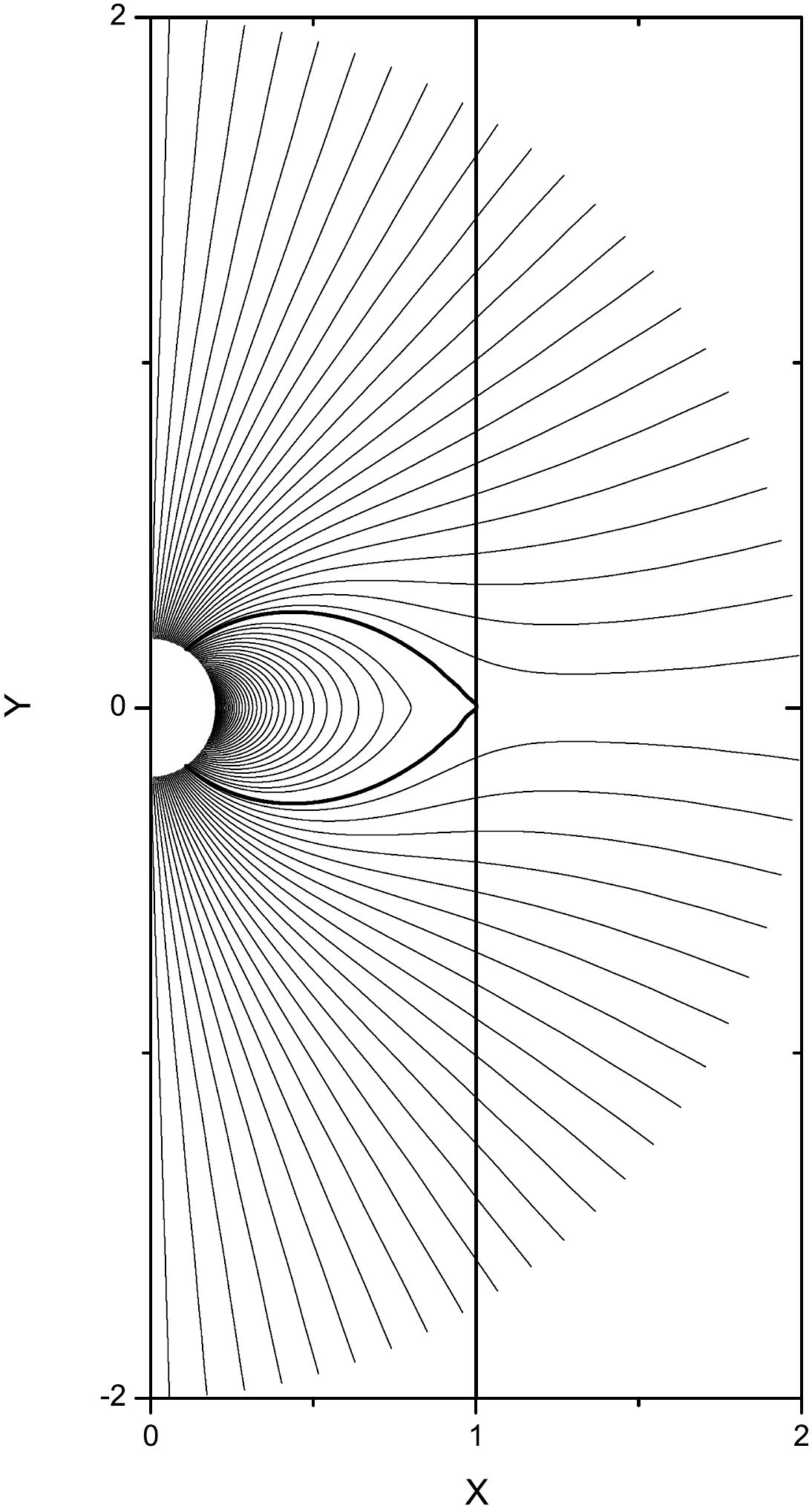, width=6cm}}
\caption{ Poloidal field lines of the aligned rotator. The vertical line is the light cylinder. The thick line is the field line that touches the light
cylinder.
\label{fig4}}
\end{figure}

\begin{figure}
\centerline{\epsfig{file=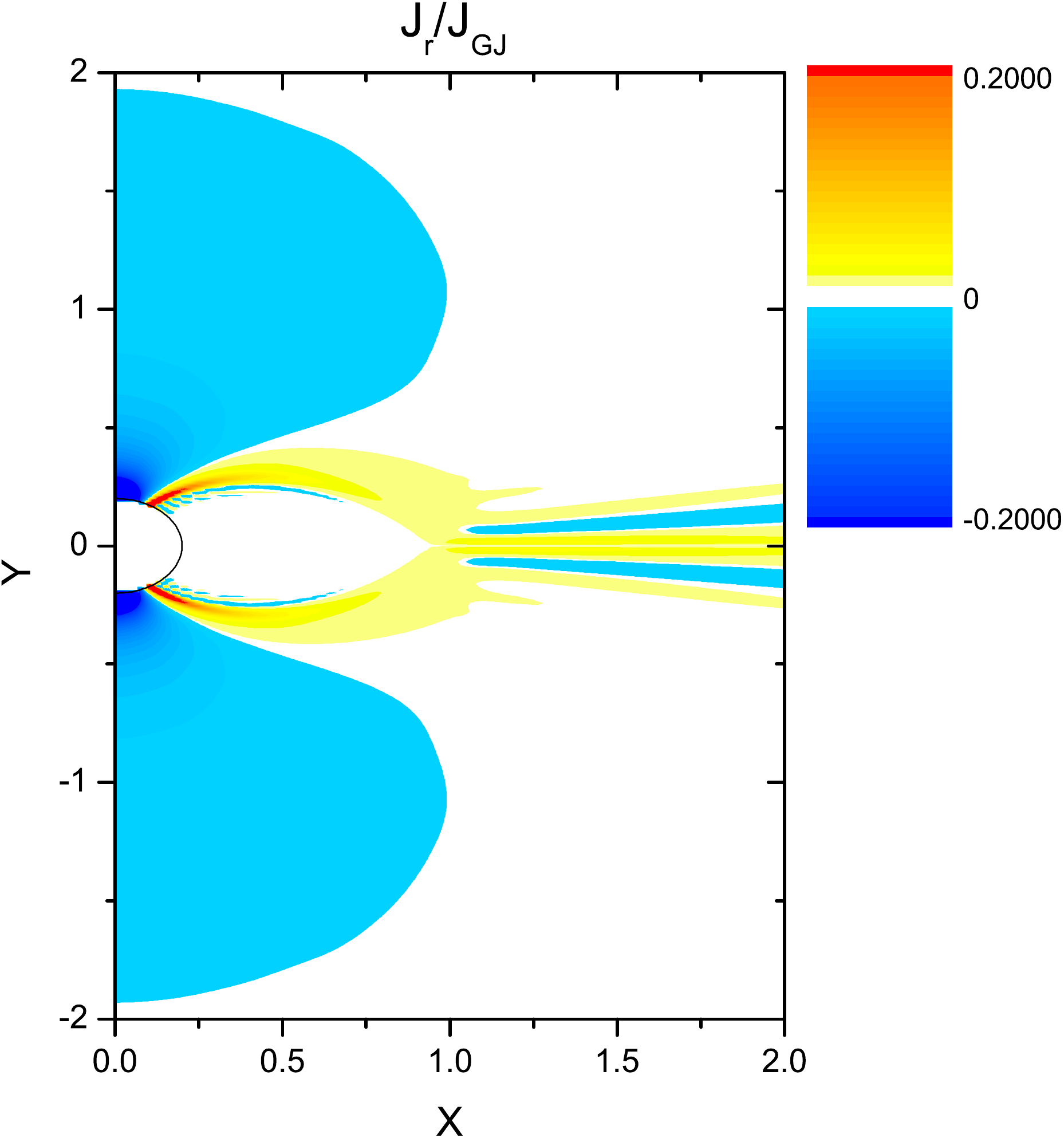, width=8cm}}
\caption{ The radial current density in the steady solution. \label{fig5}}
\end{figure}
\begin{figure}
\centerline{\epsfig{file=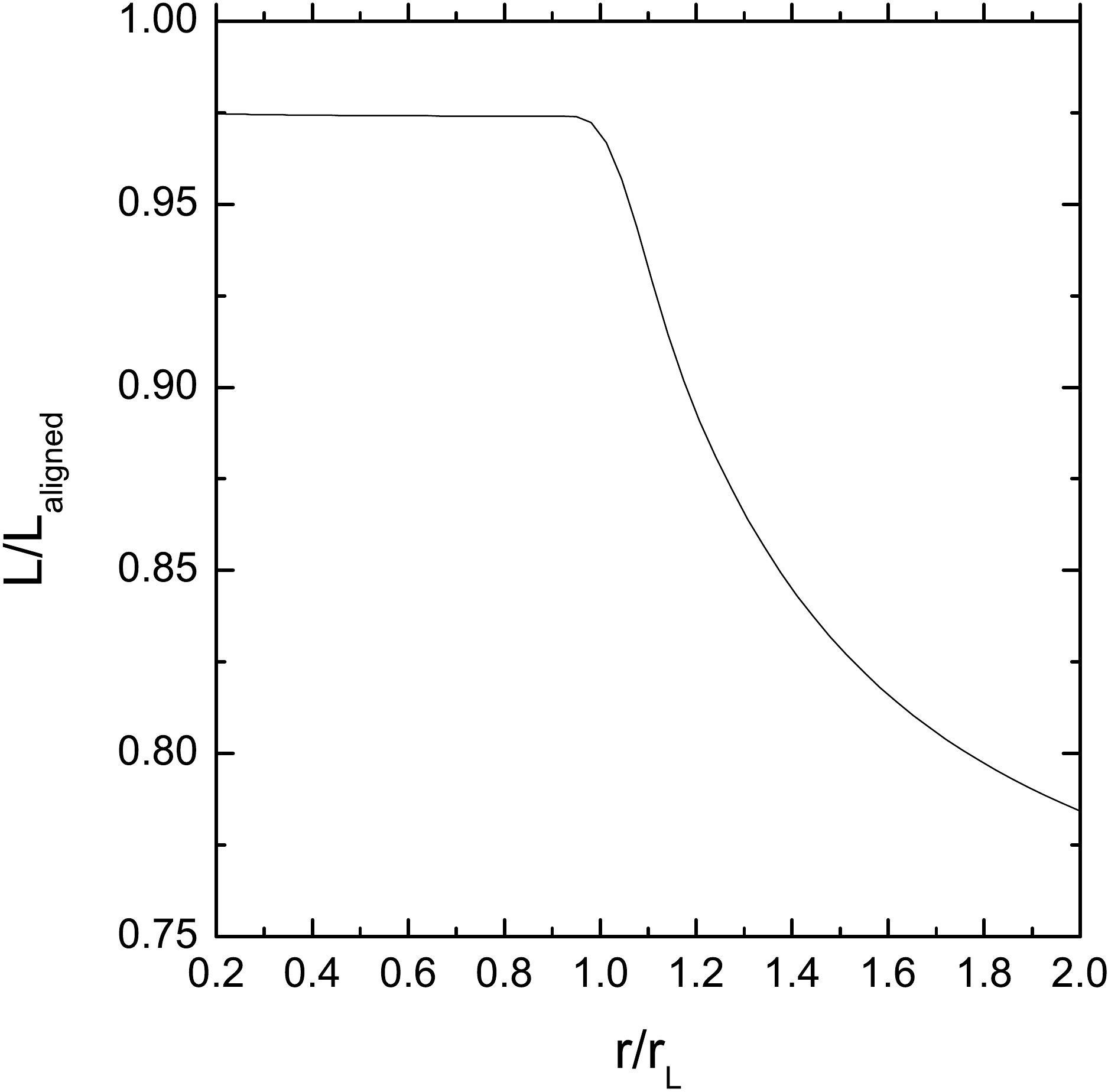, width=7cm}}
\caption{ The normalized Poynting flux $L/L_{\rm aligned}$ across the sphere of radial $r$. \label{fig6}}
\end{figure}

\begin{figure}
\centerline{\epsfig{file=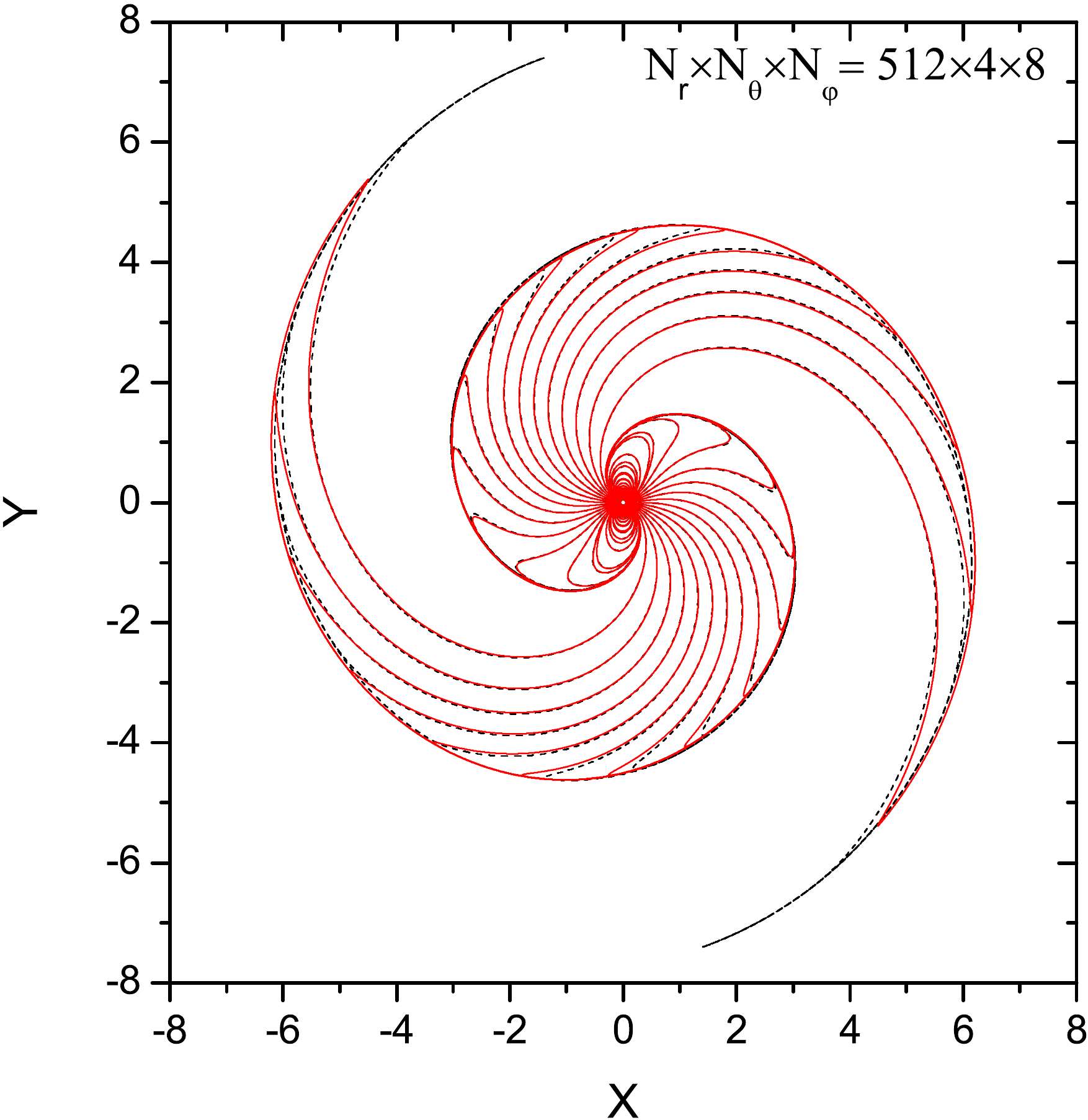, width=7cm}}
\caption{ Magnetic field lines of the perpendicular Deutsch field in the equatorial plane. The numerical and analytical solutions are shown as the black dashed and red solid lines, respectively. \label{fig7}}
\end{figure}

\section{DISCUSSION AND CONCLUSION}

In this paper, to solve a set of the time-dependent force-free equations, we have developed a pseudo-spectral code  through improving the method of \citet{pet12} by implementing an absorbing boundary layer. Two famous analytical solutions of the Deutsch vacuum dipole (Figure 1) and the Michel monopole (Figure 2) can be reproduced well by our pseudo-spectral code. Meanwhile, the time-dependent simulations of the axisymmetric force-free pulsar magnetosphere has been carried out. Our results show that (1) the current sheet in the equator plane can be resolved well, (2) a steady solution with some closed lines beyond the light cylinder is obtained because of the numerical dissipation by spectral filtering (see Figures 3), (3) the radial current density in the steady-state solution is smaller than the GJ current density (Figure 5), and (4) the Poynting flux inside the light cylinder (Figure 6) is in good consistency with the value given by \citet{spi06}.

Our code can be extended to three-dimensional (3D) geometry by including the Fourier transform in the azimuthal direction. This would allow us study the structure of the oblique pulsar magnetosphere. For the purposes of illustration, we show the 3D simulation for the perpendicular vacuum rotator in Figure 7. The figure shows that the numerical solution is in good agreement with the analytical solution. The simulation demonstrates that our code can be extended to 3D oblique rotator by including the azimuthal transformation. In our simulation, the inner boundary is set to be $R_{\star}=0.05 \, r_{\rm L}$, which corresponds to the pulsars with a period of 4 ms. In principle, we can place the stellar radius to infinitesimal size as long as the resolutions are enough high, which becomes possible for us  to study the high-energy emission from the observed millisecond pulsars with the realistic ratios of stellar $R_{\star}$ and light-cylinder $r_{\rm L}$, as the Crab pulsar. However, in the finite difference method, the code cannot work well for the low stellar radius in the Cartesian coordinate, such as $R_{\star}<0.2 \,r_{\rm L}$  \citep{kal09}. Currently, the available computational resources cannot allow us perform the very high resolution simulations. The simulations presented here were obtained on a single processor.
Therefore, we will improve the code by optimization or parallelization techniques and study the absorbing boundary layer for the oblique pulsar magnetosphere in the 3D simulation. Further, we will study more realistic pulsar magnetosphere through relaxing the force-free approximation to a resistive electrodynamics in future, as those in \citet{li12} and \citet{kal12}.

\section*{Acknowledgments}
We thank the anonymous referee for valuable comments and suggestions. We would like to thank J\'{e}r$\hat{\rm o}$me P\'{e}tri, Kyle Parfrey
and Cong Yu for some useful discussions. This work is partially supported by the National Natural Science Foundation of China (NSFC 11433004, 11103016, 11173020), the Doctoral Fund of the Ministry of Education of China (RFDP 20115301110005), and the Top Talents Programme of Yunnan Province (2015HA030).


\bibliography{refernces}


\end{document}